# The influence of changes in water content on the electrical resistivity of a natural unsaturated loess

J. A. Muñoz-Castelblanco, J. M. Pereira, P. Delage and Y. J. Cui

Ecole des Ponts ParisTech, Laboratoire Navier/CERMES, Université Paris-Est.
6-8 av. B. Pascal. F 77455 Marne la Vallee cedex 2

**ABSTRACT**

Non-destructive methods of measuring water content in soils have been extensively developed in the last decades, especially in soil science. Among these methods, the measurements based on the electrical resistivity are simple and reliable thanks to the clear relationship between the water content and the electrical resistivity of soils. In this work, a new electrical resistivity probe was developed to monitor the change in local water content in the triaxial apparatus. The probe is composed of two-pair of electrodes, and an electrical current is induced through the soil at the vicinity of the contact between the probe and the specimen. Some experimental data on the changes in resistivity with the degree of saturation were obtained in specimens of a natural unsaturated loess from Northern France. Two theoretical models of resistivity were also used to analyze the obtained data. Results are finally discussed with respect to the loess's water retention properties.

## INTRODUCTION

The measurement of water content in soils has been performed through different non-destructive techniques, including neutron scattering (Chapellier 1987, Gardner and Kirkham 1951, Kruschwitz and Yaramanci 2004) and gamma ray attenuation (Reginato and van Bavel 1964). Electromagnetic sensors have also been used, including capacity sensors (Tran *et al*. 1972), ground-penetrating radar (GPR) (Chanzy *et al*. 1996) and time domain reflectometry sensors (TDR) (Davis and Chudobiak 1975, Noborio 2001, Topp *et al.* 1980). The main concern about the use of electromagnetic methods in fine grained soils is related to the dielectric dispersion due to the water located within particle aggregations in clays (Cosenza and Tabbagh 2004).

Various authors also determined the water content from the measurement of the soil's electrical resistivity. Kalinski and Kelly (1993) presented some resistivity measurements performed on a high plasticity clay in a Miller box (Fowles 1980) and in a resistivity cell (Gorman 1988, Gupta and Hanks 1972, Rhoades *et al*. 1976, Rhoades *et al*. 1977). The two protocols differ in the cell shape (square box or circular cell) but in both methods, an electric current is induced between various rods that are placed along the perimeter of the specimen. The electric resistivity is estimated from the electric potential difference between the rods. Chen *et al.* (2007) developed a system called "2D resistance grid model" to monitor the formation of a crack in a specimen. This device, close to that developed in this work, consisted in using 6 electrodes placed in a 2D grid along a sample plane. The changes in resistivity between pairs of electrodes were related to the changes in degree of saturation and to the onset and growth of the crack. A similar device was also developed by Michot *et al.* (2001). McCarter (1984) and Fukue *et al.* (1999) used a single pair of circular electrodes that were placed on the upper and lower sides of a cylindrical soil sample. The electrodes had the same diameter as the sample.

Previous studies showed that at low water contents, the resistivity rapidly decreases with increased water content, with a rate of decrease reducing at high water contents (McCarter 1984). Other parameters like the soil density, clay fraction and soil structure also have a significant influence on the soil resistivity. Fukue *et al.* (1999) also commented about the effects of the degree of saturation, of the electrical resistivity of the pore fluid, of the shape, size and distribution of solid particles and of the ion concentration and distribution in the diffuse double layers in the clay fraction. They consider that resistivity changes become very sensitive to changes in water content when the pore water phase becomes discontinuous, particularly in dense soils at low water contents.

In this work, the relevance of electrical resistivity measurements to monitor the local water content changes of an unsaturated specimen in the triaxial apparatus is examined. To do so, the changes in electrical resistivity of a natural unsaturated loess submitted to changes in water content are investigated and the data obtained are analysed by means of two literature models. A discussion about the links between the resistivity data and the water retention properties of the loess is finally proposed.

**THE ELECTRICAL RESISTIVITY OF SOILS**

The electrical conduction of soils has been investigated in details by soil scientists (McCarter 1984, Kalinski and Kelly 1993, Fukue *et al.* 1999, Robain *et al.* 2001, Samouëlian *et al.* 2003, Chen *et al.* 2007) who considered the three phases of an unsaturated soil (solids, air and water) as parallel resistors. The soil apparent electrical resistance is a function of that of solids ($R_s$), air ($R_a$) and water ($R_w$) as follows (e.g. Chen *et al.* 2007):

$$R = \left(R_s^{-1} + R_a^{-1} + R_w^{-1}\right)^{-1} \tag{1}$$

The electrical resistance $R$ (the relation between the voltage $V$ and the current intensity $I$) is expressed in ohms (1Ω = 1V / 1A). When an electrical current is imposed to a soil specimen by a system of two electrodes (anode-cathode), the electrical resistivity $\rho$ (Ωm) is equal to the product between the electrical resistance and a geometric coefficient $k$ depending on the size and arrangement of electrodes.

In granular soils, the electrical resistance of solids is very high, whereas it is several orders of magnitude lower in clays due to the effect of the water adsorbed along the clay platelets (Samouëlian *et al.* 2003). Air is considered as an electrical insulator (Samouëlian *et al.* 2003). The apparent electrical resistivity of the soil is hence largely dependent on the amount and continuity of pore water. As an illustration, the electrical resistivity of dry sands is about $10^5$ Ωm whereas it is around 10 Ωm in saturated sands (Fukue *et al.* 1999).

Robain *et al.* (2001) related the electrical resistivity of soils to their structure, relating low and high resistivity values to macro and micro pores, respectively. The low resistivity corresponding to macro-pores is associated with a larger accessibility of water molecules to the electrical current, compared with the case of micro-pores. The electrical resistivity of a soil aggregate highly depends on its microstructure, given the significant difference between the resistivity of free water (2 to 100 Ωm) and that of silicates (almost non conductive with $\rho$ between $10^{10}$ and $10^{14}$ Ωm). As a consequence, an electrical current will almost totally flow through the pore water within the aggregates. Pore microstructure thus dictates the path of electric currents (Guéguen and Palciauskas 1994).

In the absence of any detailed information about the pore configuration, a simple empirical relation between electrical resistivity and porosity was proposed by Archie (1942) for saturated soils, in which the soil resistivity $\rho_0$ is related to that of free water $\rho_w$, as follows:

$$\frac{\rho_0}{\rho_w} = (n)^{-a} \tag{2}$$

where $n$ is the soil porosity and $-a$ a soil parameter.

Archie's second law is an extension to unsaturated soils, as follows:

$$\frac{\rho}{\rho_0} = (S_r)^{-b} \tag{3}$$

in which $S_r$ is the degree of saturation, $\rho$ the electrical resistivity of the unsaturated soil and $b$ a soil parameter (Guéguen and Palciauskas 1994).

Fukue *et al*. (1999) proposed a more sophisticated model accounting for the combined effects of serial and parallel transmission of the electric current in the three phases. They introduced a structure coefficient $F$ (identified as a relative internal length) to account for the contributions to the total electric current of both the parallel flow (related to $1 - F$ and mainly occurring in water) and the serial one (related to $F$ and mainly influenced by the insulating properties of solids and air). They proposed that the electric resistivity through a cylinder of radius $r$ be read as:

$$\frac{\rho_0}{\rho_w} = \frac{1}{n} \pi r \left( \frac{1}{1 - F_{sat}} \right) \tag{4}$$

$$\Gamma_{sat} = \frac{\rho_w}{(1 - F_{sat})} \tag{5}$$

$$\frac{\rho}{\rho_0} = \frac{1}{S_r} \left( \frac{1 - F_{sat}}{1 - F} \right) \tag{6}$$

where coefficient $F_{sat}$ corresponds to the saturated condition and $F$ varies with the degree of saturation and depends on the pore size distribution. Parameter $\Gamma_{sat}$ describes the quality of sampling in intact soils (Fukue *et al*. 1999).

**MATERIAL AND EXPERIMENTAL SETUP**

This work was performed on a natural unsaturated loess sample manually extracted near the village of Bapaume in Northern France. Loess deposits in this area were formed under periglacial conditions during the Quaternary period as the consequence of the aeolian transport of silty sediments. Loess deposits have a relative homogeneity, a low plasticity, a high porosity and an open structure that explain its susceptibility to collapse (Cui *et al*. 2004, Delage *et al*. 2005, Cui *et al*. 2007, Yang *et al*. 2008, Karam *et al*. 2009). The main geotechnical properties of the loess are summarized in Table 1. Cylindrical specimens of diameter 70 mm and height 17.5 mm were carefully trimmed from intact blocks that were extracted at different depths of a 5 m deep excavation.

A small sized electrical resistivity probe was developed to measure the electrical resistivity under water content variations. It was inspired from the concentric surface probe developed by Maryniak *et al.* (2003). The probe is composed of four circular electrodes of diameter 1.5 mm disposed in a squared-grid scheme as presented in Figure 1. Figure 2 shows how the resistivity probe is fixed on top of the soil specimen. This probe was developed to monitor water content changes in the triaxial apparatus but calibration was carried out on a simpler device with no mechanical loading. The system is composed of a precision balance to measure the water mass changes, of a metallic cylindrical mould housing the specimen and of a plastic cover disk accommodating the resistivity probe and covering the specimen to avoid any evaporation during the measurements. The plastic disk is removable in order to allow sample wetting and drying. Good contact between the probe and the soil is ensured by placing a fine wet layer of loess slurry on the probe. Ambient laboratory temperature is controlled at 20°C ± 0.5°C to avoid any parasite changes in electrical resistivity due to temperature variations (Kalinski and Kelli 1993 recommended to correct the measured electrical resistivity at temperatures above 21°C according to the procedure in ASTM G57).

The diameter of the resistivity probe is 11 mm and the distance between the electrodes is 6 mm. A hydrophobic isolating dielectric matrix made up of an epoxy resin (Araldite 2012) is used to accommodate the electrodes in order to avoid any direct current line between them. The electrodes' disposition is presented in Figure 3. The two input electrodes are connected to a voltage source of 10 V. The current passes through the soil and the signal is received by the two other output electrodes. For a circuit in parallel, the apparent electric resistance of the probe, R, reads as follows:

$$\frac{1}{R} = \frac{1}{R_{s1}} + \frac{1}{R_{s2}}; \qquad (7)$$

giving $R_s = 2R$ since $R_{s1} = R_{s2} = R_s$.

The shape of the current lines between the input and output electrodes is related to the geometry and boundary of the problem. The electrical resistivity of soil is given as:

$$\rho = \frac{R_s A_e}{L} \qquad (8)$$

where $R_s$ is the measured soil electric resistance, $A_e$ is the electrode surface and $L$ is the shorter distance between each pair of electrodes.

**EXPERIMENTAL INVESTIGATION**

Tests were performed to characterize the relationship between the changes in water content and that in soil resistivity. Six loess specimens of height 17.5 mm and diameter 70 mm were subjected to controlled wetting and drying processes while measuring the electrical resistivity. Three specimens from 1 m depth (*e* = 0.84) and two others from 3.3 m depth (*e* = 0.60) were tested. Another loess specimen from 1m depth with a void ratio of 0.72 was also tested in order to study the influence of the porosity on the resistivity response. This sample was obtained by compressing a natural sample (*e* = 0.84, *w* = 14.4%) in an oedometer down to *e* = 0.72 (axial strain rate of 1.7 µm/min).

The decreases in water content were achieved by letting the sample dry inside the oedometer cell in the laboratory for various periods of time comprised between 10 and 24 h. Wetting was achieved by carefully adding small quantities of water to the soil sample: a piece of filter paper was placed on top of the sample and water drops were uniformly distributed over the filter paper by using a syringe. To ensure water content homogeneity in both drying and wetting cases, the samples were set to rest for one day after the change in moisture conditions (in the wetting process, the filter paper piece was kept on during equilibration and the cover disk was placed to avoid evaporation). Water content changes at equilibrium were controlled by weighing with an accuracy of 1/1000 g (the filter paper was removed at each weighing operation so as to only tare the oedometer cell and the soil sample). The resistivity of the tap water used to wet the samples was measured, giving an average value of 3.5 Ωm close to that average value of 3 Ωm measured on the loess pore water. This low resistivity value compared to distilled water (more than 1000 Ωm) is related to the salt content of both tap water and pore water.

Figure 4 shows the data obtained in all the samples along both the wetting and drying paths. The relationship between the electrical resistivity and the volumetric water content $\theta$ is presented in Figure 4a in a semi-log plot. Data are also plotted with respect to the degree of saturation in Figure 4b. At each depth, a fairly good compatibility between the data from the different specimens tested is observed along both the wetting and drying paths.

The sample extracted at 1 m depth ($e = 0.84$) exhibits a soil resistivity increase from 8 Ωm to 338 Ωm when the volumetric water content decreases from 38.5 to 5.8% (gravimetric water content $w$ decreasing from 27.0 to 4.0%). The slope of the curve indicates that a reasonable estimation of the water content can be made from the electrical resistivity at volumetric water contents higher than 10.0% ($w > 7.0\%$). In a drier state ($w < 7.0\%$), the resistivity rapidly reaches values higher than 50 Ωm and the changes become too tiny to allow a good accuracy in the estimation of the water content. A strong similarity is observed between the resistivity data at $e = 0.84$ and $e = 0.72$. In general, the resistivity is slightly higher for the densest sample ($e = 0.72$), especially at degrees of saturation higher than 0.7 (see Figure 4b). However, this difference appears to be almost negligible if the data are plotted in terms of volumetric water content $\theta$.

The resistivity curve of the 3.3 m depth specimens appears to be steeper than that of the 1 m depth samples, allowing a more accurate determination of the water content that can satisfactorily be made between $\theta = 15.0\%$ and $\theta = 27.0\%$ (gravimetric water contents between $w = 8.7\%$ and $w = 16.9\%$). Note that the samples from the two depths have approximately the same index properties ($w_p = 19\%$, $w_L = 28\%$ at 1m and $w_p = 21\%$, $w_L = 30\%$ at 3.3 m depth, see Table 1). The deeper samples are denser ($e = 0.60$ at 3.3 m compared to 0.84 at 1 m), with a higher clay fraction (25% at 3.3 m compared to 16% at 1 m), and a higher initial degree of saturation ($S_r = 0.72$ at 3.3 m compared to 0.46 at 1 m). This explains the lower initial resistivity at natural state of the deeper sample (19 Ωm at 3.3 m compared to 30 Ωm at 1 m), given that a larger proportion of the pore volume is full of water. The point corresponding to the soil resistivity at natural water content of the 1m depth samples ($\rho = 30$ Ωm at $w = 14.4\%$) is also represented in Figure 4.

## ANALYSIS AND DISCUSSION OF THE RESISTIVITY DATA

For both soils, the changes in relative electrical resistivity $\rho/\rho_0$ is plotted with respect to the degree of saturation $S_r$ in a log-log graph presented in Figure 5. Data are compared to Archie's second law and to the model proposed by Fukue *et al.* (1999). The resistivity of the pore water $\rho_w$ is estimated to be equal to 3 Ωm which is in the range of 2 to 100 Ωm given for natural fresh water (Palacky 1987). The saturated soil resistivity $\rho_0$ is equal to 10.3 and 10.1 Ωm for the 1m depth samples with $e = 0.84$ and 0.72 respectively, and to 9.8 Ωm for the 3.3 m loess specimens ($e = 0.60$). These comparable resistivity values under saturated conditions show that the initial porosity has little effect.

The curves obtained from Archie's law agree reasonably well with the experimental data at degrees of saturation higher than $S_r = 0.20$ and $0.30$ for the 1 m specimens and 3.3 m specimens, respectively. Given their concave shape, Fukue's curves better correspond to the experimental data than Archie's law. $F_{sat}$ values (see Equation (6)) are 0.98 for the 1m loess and 0.99 for the 3.3 m loess, corresponding to $\Gamma_{sat}$ values of 305 $\Omega$ and 366 $\Omega$ respectively. Fukue et al. (1999) mentioned that $\Gamma_{sat}$ values near or higher than 300 $\Omega$ indicate good quality specimens, in accordance with the block sampling technique used here.

Figure 5 also shows that the resistivity is very high at the lowest degrees of saturation and very low at the highest ones. The relationship also presents an abrupt increase in resistivity at $S_r < 0.20$ for the 1m depth specimens ($w = 6\%$, $e = 0.84$ and $w = 5\%$, $e = 0.72$) and at $S_r < 0.30$ for the 3.3m depth specimens ($w = 7.4\%$). This could result from the transition between continuity and discontinuity of the water phase within the clay fraction (Fukue et al. 1999). Beyond the critical degree of saturation, the resistivity correctly fits with Archie's second law.

Using expression (6) and according to Fukue's model, the saturated (1-$F_{sat}$) and unsaturated (1-$F$) structural parameters can now be compared, as done in Figure 6 by plotting the changes in the (1-$F_{sat}$/1-$F$) with respect to the changes in volumetric water content. Both samples from 1m depth (with two different void ratios) exhibit the same trend. At volumetric water contents higher than $\theta = 7.5\%$ ($w = 5\%$ at $e = 0.84$ and $w = 4\%$ at $e = 0.72$), the ratio (1-$F_{sat}$)/(1-$F$) is fairly constant and close to unity, whereas it sharply increases at smaller water contents (below $\theta = 6\%$). The same data plotted in Figure 6(b) for the 3.3m deep samples exhibit a different trend, with an increase from 1 (at $\theta = 29.0\%$ and $w = 18\%$, a value close to the initial natural value of 17.9%) to 5 (at $\theta = 13.8\%$, $w = 7.4\%$), followed by a sharp increase at lower volumetric water contents. As mentioned above, the beginning of this increase in electric resistivity is related to the transition from continuity to discontinuity of the water phase within the clay fraction.

Figure 7a and b represent in the same graph both the relative resistivity / water content curve and the water retention curve of the samples from 1 and 3.3 m depths, respectively. The water retention curve at 1 m was taken from Muñoz-Castelblanco et al. (2011) whereas that at 3.3 m was obtained by using the filter paper method to measure the changes in suction of samples that were carefully dried and wetted using the techniques described above. The resistivity ratio $\rho/\rho_0$ (left y-axis) and suction (right y-axis) are represented in a logarithmic scale whereas the water content (upper x-axis) and the degree of saturation (lower x-axis) are presented in a linear scale. The water retention curve of the 1 m specimens exhibits a hydraulic hysteresis that is less evident in the 3.3m specimens. Figure 7 also shows that, at comparable water contents, the 1 m specimens exhibit lower suctions than the 3.3 m specimens. This is due to the probable smaller size of the predominant pores population of the denser 3.3m specimens compared to the larger pores of the more open structure of the 1 m specimens.

The resistivity and water retention curves have comparable shapes. However, there is no hysteresis effect on the resistivity curve, showing that the resistivity is only dependent on the water content, irrespective of the suction value or on the drying or wetting path followed to reach the water content considered. This confirms the interest of resitivity measurement to monitor changes in water content.

Finally, a well defined relation has been obtained between the water content and the resistivity that may be used for calibration purposes. The measurement of water content would not be as reliable in the ranges where the electrical resistivity sharply increases with water content.

According to field suction measurements on shallow loess deposits in Northern France described in Cui *et al.* (2008), in-situ suction variations were observed from 10 kPa to about 200 kPa. According to the water retention curve of loess 1m depth (Figure 7), this suction range corresponds to gravimetric water content changes from 7.5% to 22.5%. In this case, provided temperature effects are also calibrated, the probe could be used to measure in-situ water content changes under climatic effects.

**CONCLUSIONS**

The electrical resistivity of a natural unsaturated loess from Northern France was measured under various water contents. A deviation from Archie's law was observed at low degrees of saturation. This observation is linked to the discontinuity of pore water within the clay fraction of the loess at low degrees of saturation. The curve obtained using Fukue's model agrees well with experimental data at low degrees of saturation because this model accounts for the variation of a structural factor *F* with the degree of saturation, especially at water contents lower than a "critical" threshold related to the microstructure. This critical value suggests that the electrical conduction behavior can be divided into at least two regimes: the first one corresponds to the continuous state of pore-water (for higher water contents) whereas the second one corresponds to the discontinuous state of pore-water (for lower water contents).

Electrical resistivity results were discussed with reference to the water retention properties of the loess. It has been confirmed that the electrical resistivity is mainly related to the amount of water. Unlike the water retention curve, the change in resistivity during a drying-wetting cycle did not exhibit any hysteresis. This study also showed that the influence of porosity changes on the resistivity – soil moisture curve may be neglected in this type of soil. Finally, the results presented in this work support the use of resistivity probes to measure water content of unsaturated soils.

**Acknowledgements**

The present study is part of the first author PhD work. It was supported by the European Alβan Program of high level scholarships for Latin America, scholarship N° E07D402297CO, through grants to Mr. J. Muñoz. The support of the French Railways Company SNCF is also acknowledged.

**Table 1. Geotechnical properties of the Bapaume loess at two different depths.**

| Sample depth | 1 m | 3.3 m |
|---|---|---|
| Natural water content $w$ (%) | 14.0 | 17.9 |
| Natural void ratio $e$ | 0.84 | 0.60 |
| Dry unit mass $\rho_d$ (Mg/m$^3$) | 1.45 | 1.67 |
| Natural degree of saturation $S_r$ | 0.46 | 0.72 |
| Natural suction (HTC) (kPa) | 40 | 48 |
| Clay fraction (% < 2 μm) | 16 | 25 |
| Plastic limit $w_p$ | 19 | 21 |
| Liquid limit $w_l$ | 28 | 30 |
| Plasticity index $I_p$ | 9 | 9 |
| Carbonate content (%) | 6 | 5 |
| In situ total vertical stress $\sigma'_{v0}$ (kPa) | 15.47 | 35.57 |

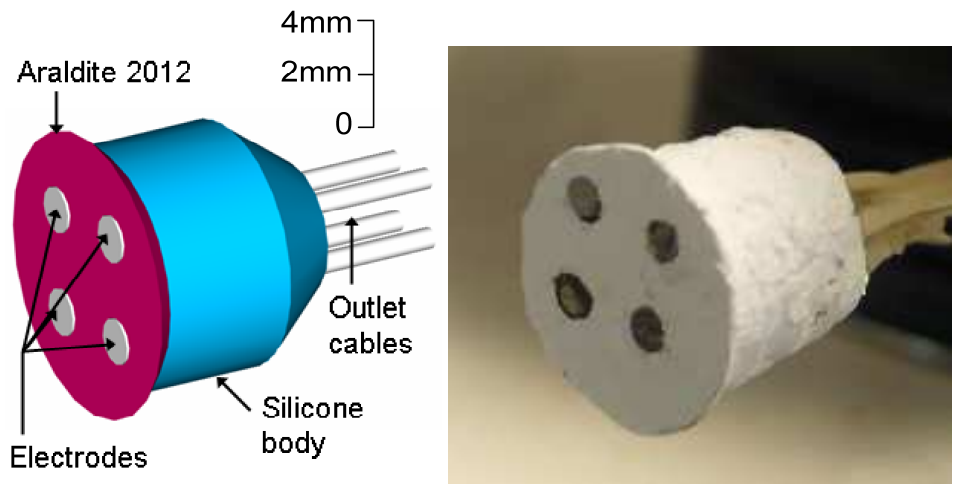

Figure 1. Home-made resistivity probe

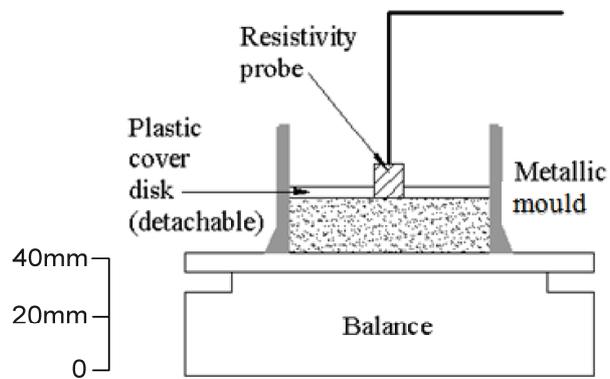

Figure 2. Experimental setup

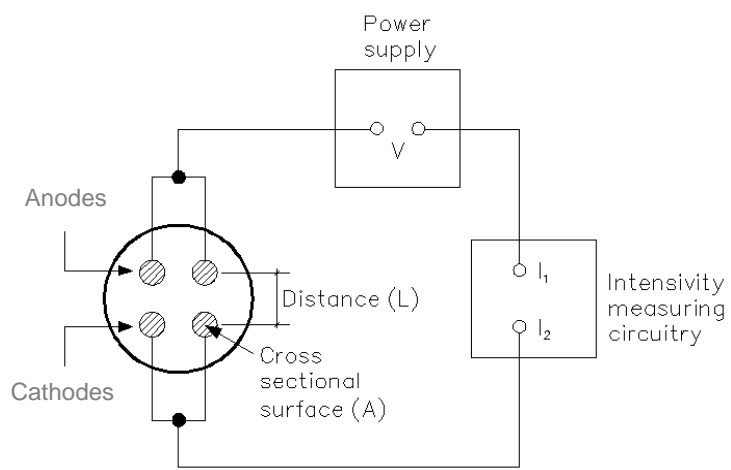

Figure 3. Electric resistivity device with four electrodes.

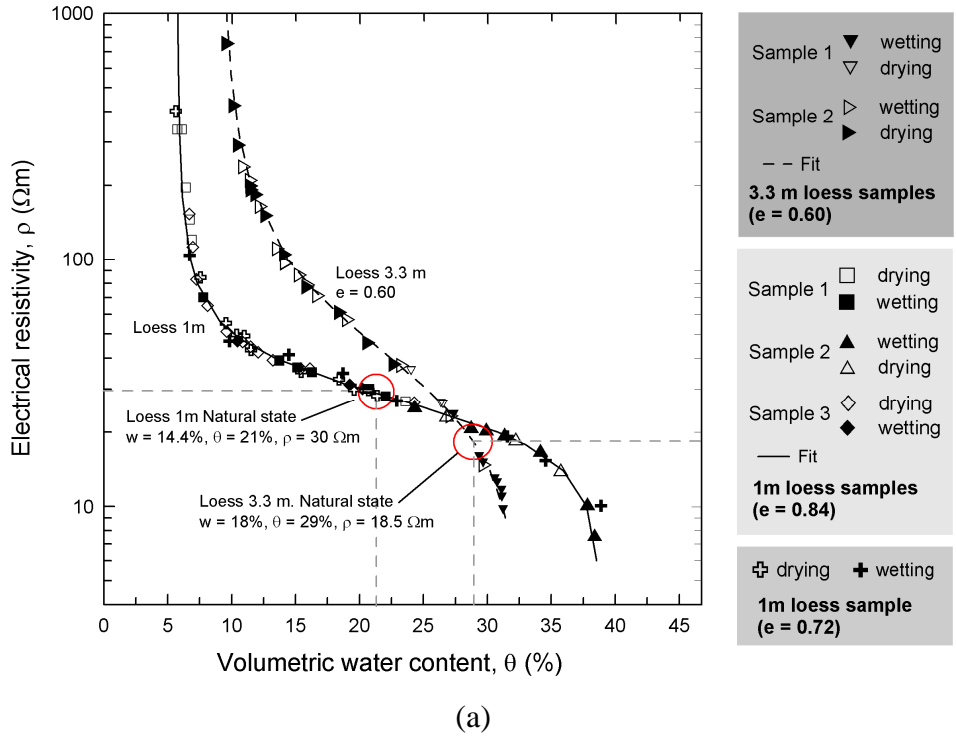

(a)

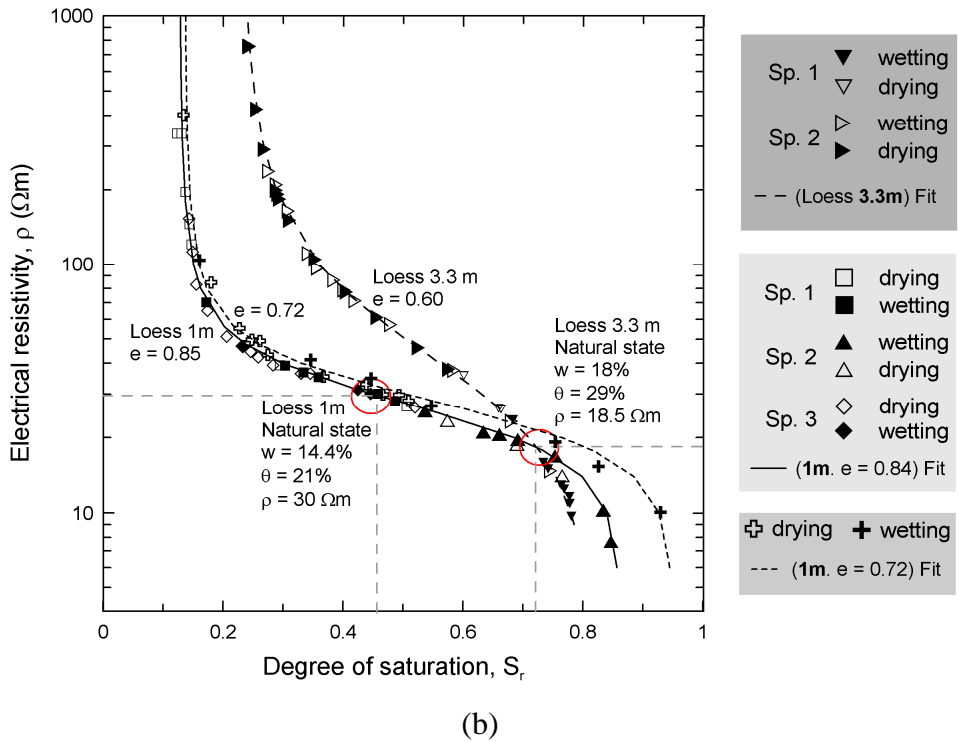

(b)

Figure 4. Resistivity data for loess at 1m depth (e = 0.84 and e = 0.72) and 3.3 m depth. Electric resistivity versus (a) volumetric water content (b) degree of saturation.

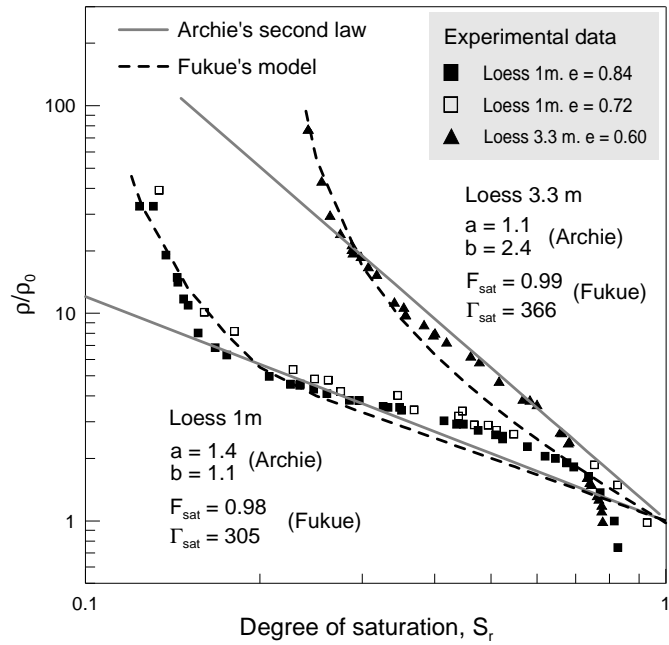

Figure 5. Resistivity data. Comparison with Archie's second law expression and with Fukue's model

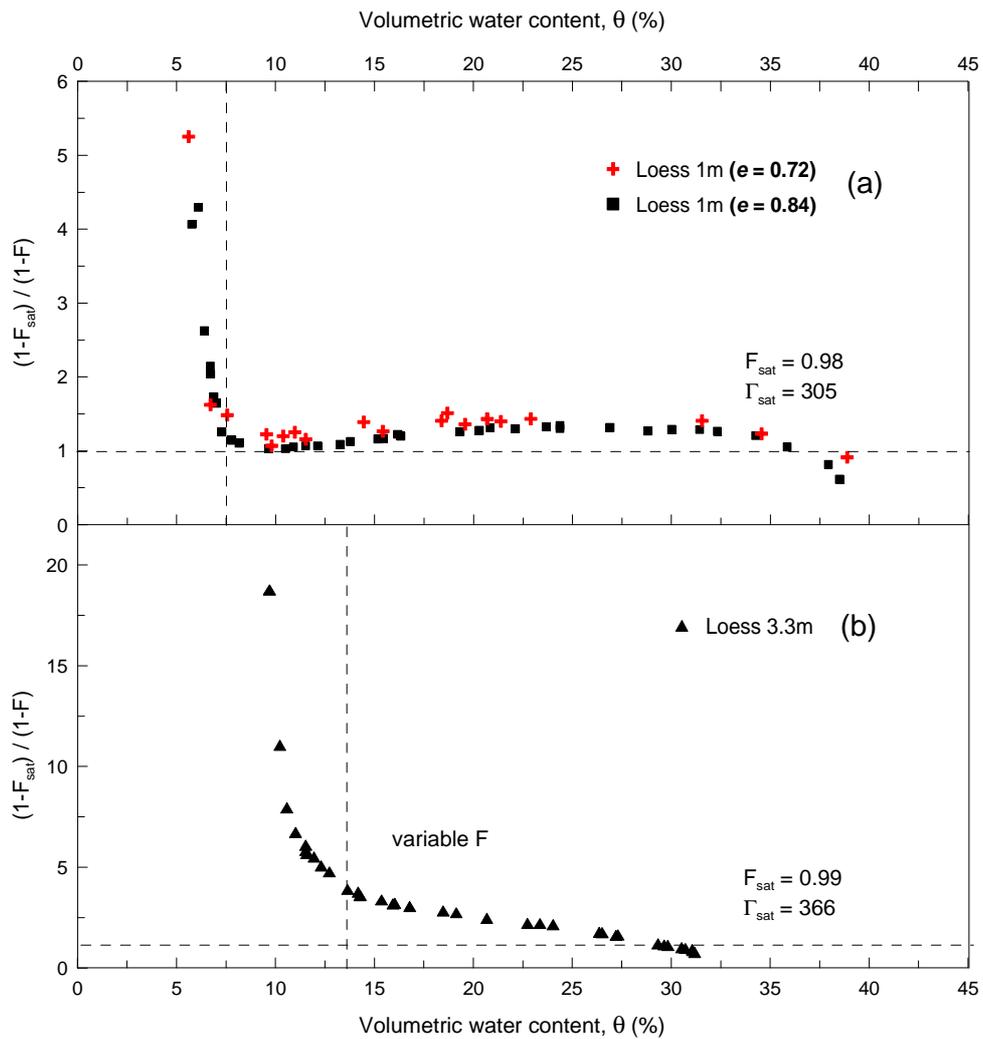

Figure 6. Comparison of the structural factors for loess specimens (Fukue's model)

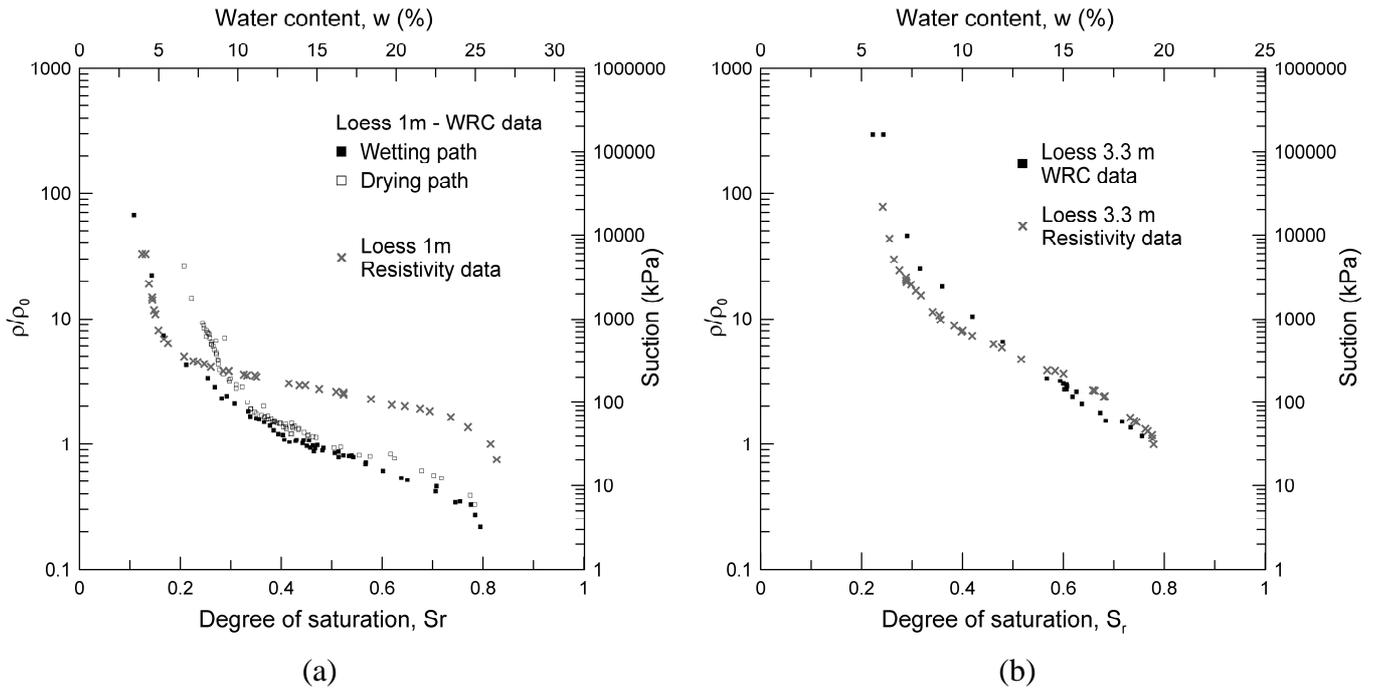

Figure 7. Comparison between water retention curves and resistivity data